\documentclass[twocolumn,showpacs,preprintnumbers,prb]{revtex4-1}
\usepackage{graphicx,bm,amsmath,amssymb}

\def\gz{\ifmmode{Z\hskip -4.8pt Z}
    \else{\hbox{$Z\hskip -4.8pt Z$}}\fi}

\newcommand{\be}{\begin{equation}}
\newcommand{\ee}{\end{equation}}
\newcommand{\bea}{\begin{eqnarray}}
\newcommand{\eea}{\end{eqnarray}}

\begin{document}

\title{Comment on ``Conductance scaling in Kondo-correlated quantum dots: 
Role of level asymmetry and charging energy''}
\author{A.~A.~Aligia}
\affiliation{Centro At\'{o}mico Bariloche and Instituto Balseiro, Comisi\'{o}n Nacional
de Energ\'{\i}a At\'{o}mica, 8400 Bariloche, Argentina}
\email{aligia@cab.cnea.gov.ar}

\begin{abstract}
In a recent work [L. Merker, S. Kirchner, E. Mu\~{n}oz, and T. A. Costi,
Phys. Rev. B \textbf{87}, 165132 (2013)], the authors compared results of numerical renormalization group
and a perturbative approach for the dependence 
on temperature $T$ and magnetic field $B$ of the conductance through a quantum dot
described by the impurity Anderson model, for small $T$ and $B$. We show that 
the equation used to extract the dependence on $B$ from NRG results is incorrect
out of the particle-hole symmetric case. As a consequence, in the Kondo regime, the correct NRG
results have a weaker dependence on $B$ and the disagreement between both approaches increase. 

\end{abstract}

\maketitle

%\date{\today }

Recent experimental studies for the conductance through one quantum dot (QD)
for low applied bias voltage $V$ and temperature $T$,\cite{grobis,scott}
stimulated further theoretical work on the 
subject.\cite{rinc,sela,rati,roura,bal,sca,ng,mu,merker} Using a Fermi liquid
approach, based on perturbation theory in $U$ (PTU), and Ward identities,
Oguri had determined exactly the scaling up to second order in $T$ and $V$
for the symmetric impurity Anderson model (SIAM) in which the energy level 
$E_{d}=U/2$.\cite{ogu1,ogu2} Further work considered the effect of higher
order contributions using different approximations, like PTU,\cite{rinc} $%
1/N $ expansion,\cite{rati} non-crossing approximation,\cite{roura} or
decoupling of equations of motion.\cite{bal} The effect of asymmetric
coupling to the left and right leads $\Gamma _{L}\neq \Gamma _{R}$, and
asymmetric drop in the bias voltage has been calculated up to second order
in $T$ and $V$ using Fermi liquid approaches, for the SIAM.\cite%
{rinc,sela,sca} The more general expression was given first by Sela and
Malecki \cite{sela} and reproduced by us using renormalized PTU.\cite{sca}
These results are exact up to terms of total second order in $V$ and $T$.

Some of these results were extended for $E_{d}\neq U/2$ using two different
approaches.\cite{sca,mu} A controversy between the authors of both works
exist.\cite{com,reply,rr} We claim that lesser and greater self energies and
Green functions in Ref. \onlinecite{mu} are incorrect. 
In turn, Mu\~{n}oz {\it et al.} \cite{reply} claim that a Ward identity is
not satisfied in Ref. \onlinecite{sca}. However, direct evaluation shows
that the Ward identity is in fact fulfilled.\cite{ng,rr}

While the conductance
can be expressed in terms of the retarded Green function only (which is by
construction correct in the SIAM), if the lesser and greater quantities are not correct 
conservation of the current is not
guaranteed when particle-hole symmetry is broken. Therefore, the results 
out of the SIAM of Mu\~{n}oz, Bolech and Kirchner \cite{mu} might be incorrect. 
However when both approaches can be
compared, for the linear term in $V$, they give the same result.\cite{com}
In any case, for more general multilevel models, for example when interference phenomena 
are important,\cite{bege,interf,ben} lesser quantities cannot be
eliminated from the conductance, and their correct evaluation becomes crucial.

Taking into account the above objections, the recent study of Merker \textit{et al.} \cite{merker} 
is certainly of interest. The authors compare the approach of of Mu\~{n}oz {\it et al.} \cite{mu} for the
temperature and magnetic field $B$ dependence of the conductance $G$, with accurate
numerical-renormalization-group (NRG) calculations at equilibrium ($V=0$).  
For the dependence on $B$, the authors combine NRG
results for the total occupation of the localized level $n_{d}=n_{d\uparrow
}+n_{d\downarrow }$ with the Friedel sum rule for finite $B$ \cite{lan,lady}

\begin{equation}
\rho _{\sigma }(0,B)=\frac{\sin ^{2}(\pi n_{d\sigma })}{\pi \Delta },
\label{fri}
\end{equation}%
which relates the spectral density of the localized level for a given spin $%
\rho _{\sigma }(\omega ,B)$ at the Fermi level $\omega =0$ with the
corresponding occupancy. Since the conductance for each spin $G_{\sigma }(B)
$ at $T=0$ is proportional to $\rho _{\sigma }(0,B)$, expanding $n_{d\sigma }
$ up to second order in $B$ and replacing in Eq. (\ref{fri}) one obtains the
corresponding expansion in the total conductance $G=G_{\uparrow}+G_{\downarrow}$. Specifically 

\begin{equation}
n_{d\sigma }(B)=\frac{n_{d}}{2}+\frac{\chi B}{g\mu _{B}}\sigma +\frac{%
\partial ^{2}n_{d}}{\partial B^{2}}\frac{B^{2}}{4} +O(B^3),  \label{nds}
\end{equation}
where $\chi $ is the magnetic susceptibility, $\sigma =1$ (-1) for spin up
(down) and the quantities in the second member except $B$ are evaluated at $B=0$. 

The last term is missed in Ref. \onlinecite{merker}. While this term
vanishes for the SIAM, because $n_{d}=1$ there as a consequence of
electron-hole symmetry, it becomes increasingly important out of the SIAM,
for which the perturbative approach of Ref. \onlinecite{mu} was developed.
In this work we examine the effects of this term. An important consequence
is that the results presented in Ref. \onlinecite{merker} (Fig. 8 for
example) as coming from NRG are misleading, because one expects that the are
highly accurate, but since they were obtained indirectly neglecting the
last term in Eq. (\ref{nds}), they should be corrected. We also show that
inclusion of this term increases the disagreement with the perturbative
approach of Ref. \onlinecite{mu} out of the SIAM in the Kondo regime.

Replacing Eq. (\ref{nds}) in Eq. (\ref{fri}) one obtains up to order $B^{2}$

\begin{eqnarray}
\frac{G_{\sigma }(B)}{G_{\sigma }(0)} &=&\frac{\rho _{\sigma }(0,B)}{\rho
_{\sigma }(0,0)}=1+c\frac{2\pi \chi B}{g\mu _{B}}\sigma   \notag \\
&&+\left( c^{2}-1\right) \left( \frac{\pi \chi B}{g\mu _{B}}\right) ^{2}+c%
\frac{\pi }{2}\frac{\partial ^{2}n_{d}}{\partial B^{2}}B^{2},  \label{rgs} \\
c &=&\cot \left( \frac{\pi n_{d}}{2}\right) .  \label{cott}
\end{eqnarray}%
Adding both spins, and defining $c_{B}$ and $T_{0}$ by  \cite{merker} 

\begin{eqnarray}
\frac{G(B)}{G(0)} &=&1-c_{B}\left( \frac{g\mu _{B}B}{T_{0}}\right) ^{2},
\label{gb} \\
\chi  &=&\frac{\left( g\mu _{B}\right) ^{2}}{4T_{0}},  \label{t0}
\end{eqnarray}%
one obtains

\begin{equation}
c_{B}=\frac{\pi ^{2}}{16}(1-c^{2})-c\frac{\pi }{2}\left( \frac{T_{0}}{g\mu
_{B}}\right) ^{2}\frac{\partial ^{2}n_{d}}{\partial B^{2}}.  \label{cb}
\end{equation}%
For $n_{d}<1$, $c>0$. In addition, $\partial ^{2}n_{d}/\partial B^{2}$ is
also positive, as shown by exact  Bethe ansatz results.\cite{oki} This means
that the last term of Eq. (\ref{cb}), missed in Ref. \onlinecite{merker} has
the effect of decreasing the results for $c_B$ reported as NRG ones in that work
(Figs. 6 and 8). This in turn means that in the Kondo regime ($-E_{d}\gg
\Delta $ and $E_{d}+U\gg \Delta $) the disagreement between NRG and the  the
perturbative approach of Ref. \onlinecite{mu} increases (Fig. 8 of Ref. %
\onlinecite{merker}). Only well inside the intermediate valence and weak
coupling regime $-0.75<E_{d}/\Delta <0$, $U/\Delta <1.5$, the comparison
might be good.

To estimate the effect of the correction, we have calculated $c_{B}$ for $%
U\rightarrow \infty $ in the slave-boson mean-field approximation (SBMFA).
This approach fulfills Fermi liquid properties [like Eq. (\ref{fri})] and is
expected to be semiquantitatively valid at low energies. In particular for
large $N$ and low temperatures it compares very well with exact results.\cite%
{newns} In the SBMFA, the solution of the Anderson model at $T=0$ reduces to
the self consistent solution of the following two equations for the the
Lagrange multiplier $\lambda $ and the width of the quasiparticle spectral
density $\tilde{\Delta}$  \cite{lady}

\begin{eqnarray}
\frac{\lambda }{\Delta } &=&-\frac{1}{2\pi }\sum\limits_{\sigma }\ln \left( 
\frac{\epsilon _{\sigma }^{2}+\tilde{\Delta}^{2}}{W^{2}}\right) ,  \notag \\
\frac{\tilde{\Delta}}{\Delta } &=&1-\sum\limits_{\sigma }n_{d\sigma },
\label{self}
\end{eqnarray}%
where

\begin{eqnarray}
\epsilon _{\sigma } &=&E_{d}+\lambda -\sigma g\mu _{B}B/2,  \label{aux} \\
n_{d\sigma } &=&\frac{1}{\pi }\arctan \left( \frac{\tilde{\Delta}}{\epsilon
_{\sigma }}\right) ,  \notag
\end{eqnarray}%
and $-W$ is the bottom of the conduction band assumed constant. 

\begin{figure}[h!]
\includegraphics[width=7cm]{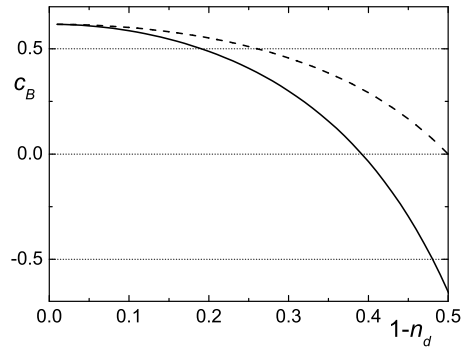}
\caption{Full line: coefficient of the magnetic field dependence of the conductance
[see Eq. (\ref{gb})]. Dashed line: the same including only the first term in 
Eq. (\ref{cb}).}
\label{cbf}
\end{figure}

After solving the problem for $B=0$, the derivatives with respect to $B$ are
obtained solving a system of linear equations, obtained differentiating Eqs.
(\ref{self}) and (\ref{aux}). The resulting $c_{B}$ is represented in Fig. 1
as a function of the occupation and compared with the result of the first
term of Eq. (\ref{cb}), which corresponds to that used in Ref. \onlinecite{merker}. 
We have chosen $W=50\Delta $. With this choice $%
n_{d}=0.99$ for $E_{d}=-5.42\Delta $ and $n_{d}=0.5$ for $E_{d}=$ $%
-2.21\Delta $. As expected, both results coincide for $n_{d}\rightarrow 1$
and the first term of  Eq. (\ref{cb}) changes sign for $n_{d}=0.5$. Instead,
the correct result changes sign for $n_{d}\simeq 0.61$, corresponding to 
$E_{d}\simeq -2.7\Delta $, and decreases strongly to negative values as $1-n_d$ (or $E_d$) 
is further increased, moving to the intermediate valence region.

In the Kondo regime, the perturbative approach of Ref. \onlinecite{mu} gives vales of $c_B$ which 
lie above those given by the first term of Eq. (\ref{cb}) (which would correspond to the dashed line 
of Fig. 1 for 
large $U$).\cite{merker} This fact and the disagreement with the temperature dependence of $G$
suggest that the approach of Mu\~{n}oz, Bolech and Kirchner \cite{mu}, at least in its present form, 
fails to correctly extend the results for the SIAM for general values of $E_d$ in the Kondo regime.

The author is partially supported by CONICET. 
This work was sponsored by PIP 112-200801-01821 of CONICET, 
and PICT 2010-1060 of the ANPCyT, Argentina.


\begin{thebibliography}{99}

\bibitem{grobis} M. Grobis, I. G. Rau, R. M. Potok, H. Shtrikman, and D.
Goldhaber-Gordon, Phys. Rev. Lett. \textbf{100}, 246601 (2008).

\bibitem{scott} G. D. Scott, Z. K. Keane, J. W. Ciszek, J. M. Tour, and D.
Natelson, Phys. Rev. B \textbf{79}, 165413 (2009).

\bibitem{rinc} J. Rinc\'{o}n, A. A. Aligia, and K. Hallberg, Phys. Rev. B 
\textbf{79}, 121301(R) (2009); arXiv:0901.4326.

\bibitem{sela} E. Sela and J. Malecki, Phys. Rev. B \textbf{80}, 233103
(2009).

\bibitem{rati} Z. Ratiani and A. Mitra, Phys. Rev. B \textbf{79}, 245111
(2009).

\bibitem{roura} P. Roura-Bas, Phys. Rev. B \textbf{81}, 155327 (2010).

\bibitem{bal} C. A. Balseiro, G. Usaj, and M.\thinspace J. S\'{a}nchez, J.
Phys. Condens. Matter \textbf{22}, 425602 (2010).

\bibitem{sca} A. A. Aligia, J. Phys. Condens. Matter \textbf{24}, 015306
(2012).

\bibitem{ng} A. A. Aligia, Phys. Rev. B \textbf{89}, 125405 (2014)

\bibitem{mu} E. Mu\~{n}oz, C. J. Bolech, and S. Kirchner, Phys. Rev. Lett. 
\textbf{110}, 016601 (2013).

\bibitem{merker} L. Merker, S. Kirchner, E. Mu\~{n}oz, and T. A. Costi,
Phys. Rev. B \textbf{87}, 165132 (2013).

\bibitem{ogu1} A. Oguri, Phys. Rev. B \textbf{64}, 153305 (2001).

\bibitem{ogu2} A. Oguri, J. Phys. Soc. Jpn. \textbf{74}, 110 (2005).

\bibitem{com} A. A. Aligia, Phys. Rev. Lett. \textbf{111}, 089701 (2013).

\bibitem{reply} E. Mu\~{n}oz, C. J. Bolech, and S. Kirchner, Phys. Rev.
Lett. \textbf{111}, 089702 (2013).

\bibitem{rr} A. A. Aligia, arXiv:1310.8324

\bibitem{bege} G. Begemann, D. Darau, A. Donarini, M. Grifoni, 
Phys. Rev. B \textbf{77}, 201406(R) (2008); \textbf{78}, 089901(E) 
(2008).

\bibitem{interf} P. Roura-Bas, L. Tosi, A. A. Aligia, and K. Hallberg, Phys.
Rev. B \textbf{84}, 073406 (2011).

\bibitem{ben} L. Tosi, P. Roura-Bas, and A. A. Aligia, J. Phys. Condens.
Matter \textbf{24}, 365301 (2012).

\bibitem{lan} D. C. Langreth, Phys. Rev. \textbf{150}, 516 (1966).

\bibitem{lady} A. A. Aligia and L. A. Salguero, Phys. Rev. B \textbf{70},
075307 (2004).

\bibitem{oki} A. Okiji and N. Kawakami, J. Phys. Soc. Jpn. \textbf{51},
3192 (1982).

\bibitem{newns} D. M. Newns and N. Read, Adv. Phys. \textbf{36}, 799 (1987).

\end{thebibliography}
\end{document}